\documentclass[twocolumn,showpacs]{revtex4}
\usepackage{graphicx}
\usepackage{dcolumn}
\usepackage{bm}
\usepackage{amsmath}
\usepackage{amssymb}
\usepackage{color}

\begin{document}

\title{Quantum radiation from superluminal refractive index perturbations.}

\author{F. Belgiorno$^{1}$, S.L. Cacciatori$^{2}$, G. Ortenzi$^{3}$, V.G. Sala$^{2}$, D. Faccio$^{4}$}

\address{$^1$Dipartimento di Matematica, Universit\`a di Milano, Via Saldini 50, IT-20133 Milano, Italy\\
$^2$INFN \& Department of Physics and Mathematics, Universit\`a dell'Insubria, Via Valleggio 11, IT-22100 Como, Italy\\
$^3$Dipartimento di Matematica ed Applicazioni, Universit\`a degli Studi di
Milano Bicocca, via Cozzi 53, IT-20125 Milano, Italy\\
$^4$CNISM and Department of Physics and Mathematics, Universit\`a dell'Insubria, Via Valleggio 11, IT-22100 
Como, Italy
}
\date{\today}

\begin{abstract}
We analyze in detail photon production induced by a superluminal refractive index perturbation 
in realistic experimental operating conditions. The interaction between the refractive index 
perturbation and the quantum vacuum fluctuations of the electromagnetic field leads to the 
production of photon pairs. 
\end{abstract}

\maketitle


In the last sixty years  a significant amount of research has been dedicated to pair-production 
in external fields or in the presence of time-varying boundary conditions, 
from the renowned analysis by Schwinger to the Hawking effect  
\cite{schwinger,hawking}. 
These effects are still waiting experimental observation although advancing technologies and the increasing number of novel physical systems in which these effects are proposed, leave hope for successful experiments in the near future. In this framework we shall focus 
our attention on pair creation of photons in a novel setting, i.e. in the presence of a moving 
 refraction index perturbation. \\
\indent  
Recent advances in different fields such as analogue gravity \cite{philbin,faccio} and the generation
 of superluminal light pulses either in optical fibers or in free space using pulse shaping techniques, 
demand a detailed quantum analysis of the pulse propagation characteristics in the presence of nonlinear 
media. In particular, intense light pulses will induce through the optical nonlinear Kerr effect, a refractive 
index perturbation that travels at the same speed, $v$, of the light pulse. Examples of superluminal light pulses
 are Bessel-pulses \cite{milchberg,bonaretti,saari_bessel} and filament pulses under appropriate 
operating conditions \cite{air-fil,russian-OC}. Therefore the refractive index perturbation associated 
to such pulses will also travel superluminally.\\   
\indent   
Herein, we propose a mechanism for photon pair-production which exhibits
the following peculiar aspects: a) the phenomenon we shall describe is associated to the
propagation of a dielectric perturbation induced by a laser pulse. Only the perturbation is in movement while the medium,
 often treated in this context in terms of an electric charge or dipole, stands still. b) We find a  condition which implies that there is no pair creation unless 
the pulse velocity satisfies $v\geq \frac{c}{n_0}$, where $n_0$ is the uniform and constant 
background refraction index. The presence of a threshold velocity is the most peculiar and relevant 
feature, which distinguishes our mechanism for example from the dynamical Casimir effect (DCE). 
c) On the other hand, the phenomenon we present is distinguishable from other superluminal propagation phenomena (\cite{ginz-book,ginzburg}) such as the Cerenkov effect. We give quantitative estimates of the expected number of photons and show that these hold promise to be detected in realistic 
experimental operating conditions.\\

\noindent \underline{\sl Model}: 
We ground our model on the following hypothesis: The nonlinearity of the dielectric medium is 
relevant only as far as it allows to obtain a refractive index perturbation
traveling with velocity $v>c/n$. Electromagnetic field quantization is then carried out as if the 
dielectric medium were linear and also, in a zero-order approximation, non dispersive. 
Then we can refer to the perturbative approach introduced by Sch\"utzhold et al. in Ref.~\cite{soff}. 
In particular, we consider the interaction representation for the 
electromagnetic field in presence of a dielectric constant $\epsilon (\vec{x},t)$ depending on space and time 
(see below); we have also a background uniform and constant value $\epsilon_b=n_0^2$ of the dielectric constant 
and our disturbance is $\epsilon (\vec{x},t)-\epsilon_b$. Then we define our interaction Hamiltonian density 
(cf. \cite{soff}):
\begin{eqnarray}\label{hint}
{\mathcal H}_I = \xi \Pi^2\hspace{0.2cm}\textrm{with}\hspace{0.2cm} \xi := \frac 12 \left( \frac{1}{\epsilon (\vec{x},t)}-\frac{1}{\epsilon_b} \right),
\end{eqnarray}
where $\vec{\Pi}=\vec{D}(\vec{x},t)$ is the canonical momentum and which is valid for the case of a medium at rest \footnote{We put $\hbar =\epsilon_0 =\mu_0 =1$. Cf. \cite{soff}}. We note that this model differs from approaches that use mode-matching between inside and outside regions of the refractive index variation, e.g. \cite{belgiornoPRL} and which are intrinsically non-perturbative..
\\

\begin{figure}[t]
\includegraphics[angle=0,width=8cm]{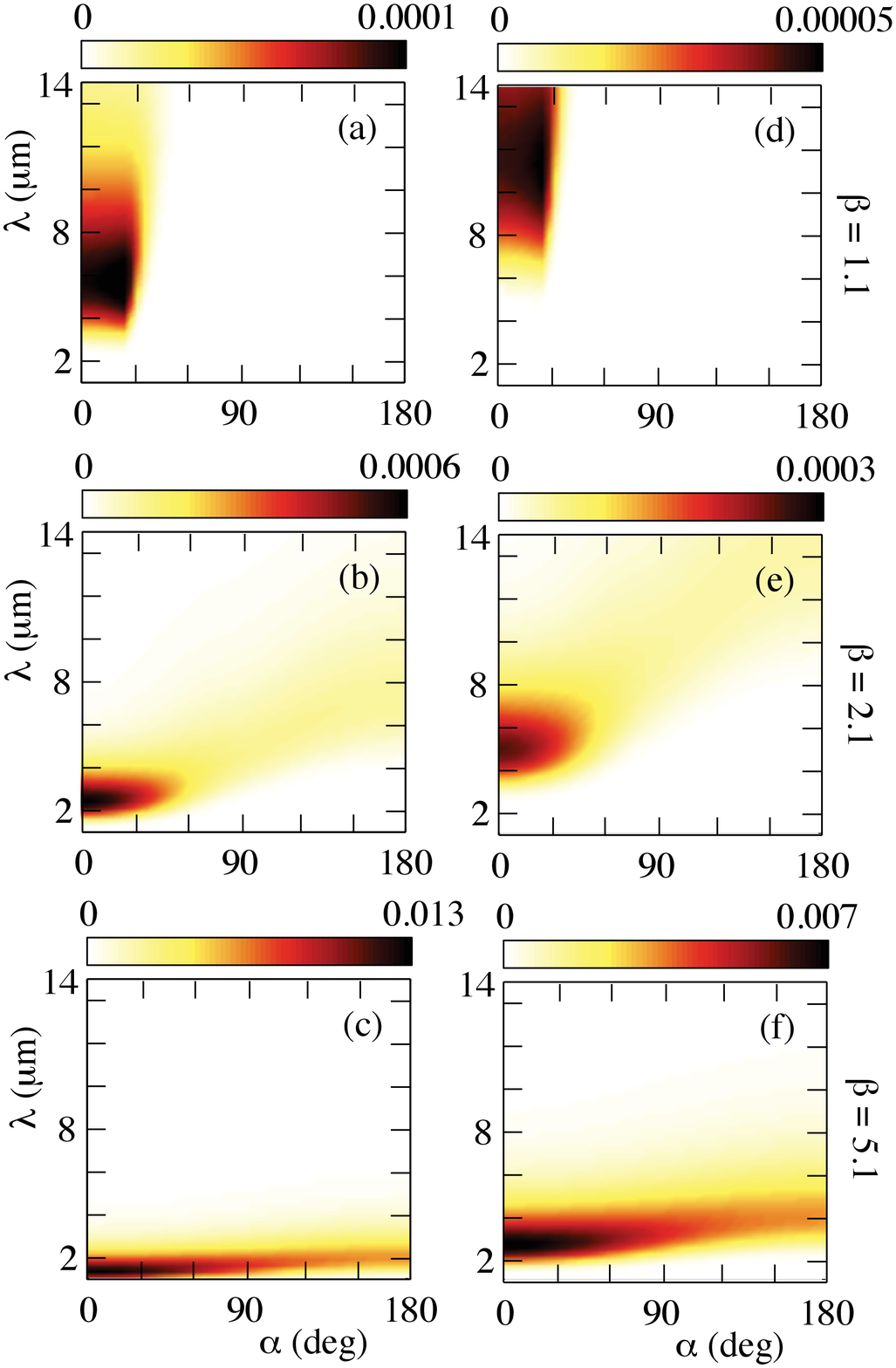} 
	\caption{ \label{fig:fig1} (in color online) Calculated photon number density from Eq.~(\ref{eq:dNdomega}) as function of emission angle, $\alpha$ and wavelength, $\lambda$. In (a), (b) and (c) $\sigma = 1$ $\mu$m and $\beta=n_0v/c=1.1$, 2.1 and 5.1 respectively.  In (d), (e) and (f) $\sigma = 2$ $\mu$m and $\beta=n_0v/c=1.1$, 2.1 and 5.1 respectively.  }
	\end{figure}

\noindent \underline{\emph{Pair-production.}}
The amplitude $A_{\{\vec{k},\mu; \vec{k}',\mu'\}}$ 
for the creation of a photon pair labeled by $\{\vec{k},\mu; \vec{k}',\mu'\}$, where 
$\mu$ is a polarization state label, is
\begin{eqnarray}
A_{\{\vec{k},\mu; \vec{k}',\mu'\}}=\langle\{\vec{k},\mu; \vec{k}',\mu'\} |S| 0 \rangle,
\end{eqnarray}
where $S$ is the S-matrix, which at the first order is given by
\begin{eqnarray}
S \simeq {\mathbb{I}} -i \int d^4x {\mathcal{H}}_I (x).
\end{eqnarray}
It may then be shown that
\begin{eqnarray}\label{pair}
|A_{\{\vec{k},\mu; \vec{k}',\mu'\}}|^2 =
\frac{\omega_k \omega_{k'}}{V^2} |\tilde{\xi}(\underline{k}+\underline{k}')|^2
(\vec{e}_{\vec{k},\mu}\cdot \vec{e}_{\vec{k'},\mu'})^2,
\label{amplitude}
\end{eqnarray}
where $\vec{e}_{\vec{k},\mu}$ are the polarization vectors and 
the tilde indicates the four dimensional Fourier transform: 
$\tilde{g}\left(\vec{k},\frac{c k}{n_0} \right)=\int dt dx dy dz\; g(\vec{x},t) 
e^{i \vec{k}\cdot \vec{x} -i \frac{c k}{n_0} t}$.
In order to find the mean number of created photons labeled by $\vec{k},\mu$, at the
same order, we have to consider the following formula:
\begin{eqnarray}
N_{\vec{k},\mu}=\sum_{\vec{k}',\mu'} |\langle \{\vec{k},\mu; \vec{k}',\mu'\} |S| 0 \rangle|^2.
\end{eqnarray}
Then we find (cf. also \cite{soff}):
\begin{eqnarray}
N_{\vec{k} \mu} = \sum_{\vec{k}'} \frac{\omega_{\vec{k}} \omega_{\vec{k'}}}{V^2} |\tilde{\xi}
(\underline{k}+\underline{k}')|^2 
[1-(\vec{e}_{\vec{k}'} \cdot \vec{e}_{\vec{k} \mu})^2 ],
\label{number}
\end{eqnarray}
where $\underline{k}=(\vec{k},\omega_{\vec{k}})=(\vec{k},{c k}/{n_0})$, with $k=|\vec{k}|$, and 
$\vec{e}_{\vec{k}}=\vec{k}/k$. 
A key-point is the following: in our case the spacetime dependence of $\xi$ takes the form 
${\xi (t,x,y,z)}\equiv \xi (x-vt,y,z)$ 
so that using the variables $u=x -v t$ and $w=x +v t$ the Fourier transform 
becomes
\begin{eqnarray}
\tilde\xi&=& 2\pi\frac{1}{v} \delta \left(k_x - \frac{c k}{v n_0}\right)\times \nonumber \\
&\times&\int du dy dz \xi (u,y,z) e^{i \frac 12 (k_x + \frac{c k}{n_0}) u}
e^{i (k_y y + k_z z)}.
\end{eqnarray}
The above result is very meaningful from a physical point of view. The dispersion relation 
$k = \sqrt{k_x^2+k_y^2+k_z^2}$ allows us to find the support of $\delta ( k_x - \frac{c k}{v n_0})$, 
which is non-zero only for $v\geq {c}/{n_0} $.  This is precisely the condition of ``superluminality'' to which we refer in relation to the present work. Clearly, there is nothing that should induce any concerns, e.g. regarding causality, in this superluminality condition and details on experimental settings in which it is reached are given below.
Note also that this condition arises for a generic $\xi (x-vt,y,z)$ as far as 
a perturbative approach is allowed. 
However, the number of photons does depend on the actual shape of the pulse.\\ 
\indent We now consider a Gaussian behavior for the refraction index 
$n^2=\epsilon (x,t,y,z)$:
\begin{eqnarray}
n^2 = n_0^2 + 2n_0 \eta e^{-[(x-vt)^2 +y^2 +z^2]/2\sigma^2}
\end{eqnarray}
which is suggested by the specific form of the pulse we can deal with in experiments. 
For $\eta\ll n_0$, which is respected in typical experimental settings as discussed below, we obtain
\begin{eqnarray}
{\mathcal H}_{I} \sim -\frac{\eta}{n_0^3} e^{-[(x-vt)^2 +y^2 +z^2]/2\sigma^2} \Pi^2.
\end{eqnarray}
Then, the number of created particles is given by the integral:
\begin{widetext}
\begin{eqnarray}
\label{Nkmu}
N_{\vec k, \mu}= 2^5 \sigma^6 \frac {\pi^2}{v^2} \frac {\eta^2}{n_0^6}  \int d\vec{k}' 
\frac{\omega_{\vec{k}} \omega_{\vec{k'}}}{V}
e^{-{\sigma^2} (|\vec{k}+\vec{k'}|^2)} \left[\delta ((k+k')_x - \frac{c}{v n_0} (|\vec{k}|+|\vec{k}'|) )\right]^2
[1-(\vec{e}_{\vec{k}'} \cdot \vec{e}_{\vec{k} \mu})^2 ].
\end{eqnarray}
\end{widetext}
The angular distribution of photons emitted in the range $k-k+dk$ is
\begin{eqnarray} \label{eq:dNdomega}
\frac {dN}{d\Omega}=\frac {2\eta^2 L \sigma^3}{\pi^2 \beta^2 n_0^6} k^3 dk \beta \gamma^2 I(\beta, \sigma, \vec k),
\end{eqnarray}
where $\beta= n_0 v/c$, $\gamma^2=1/(\beta^2-1)$ and
\begin{eqnarray}
&&I(\beta, \sigma, \vec k)=\int_0^\infty dr \int_0^{2\pi} d\theta
r\sqrt{ r^2 + f(r)^2} \cr
&&\times \; e^{-(\sigma^2 k_\perp^2+r^2 +2r k_\perp \sigma \cos \theta +(\sigma k_x+ f(r))^2)} \cr 
 && \times  \left[\frac {(k_\perp r \cos\theta +k_x f(r))^2}{k^2 (r^2+ f(r)^2)}\right] \cr
&&\times  \left[\beta+\frac {\gamma^2 \sigma (k-\beta k_x)}{\sqrt {\gamma^4 \sigma^2 (k-\beta k_x)^2+\gamma^2 r^2}}\right],
\end{eqnarray}
with $f(r)=\beta\gamma^2 \sigma(k-\beta k_x) +\sqrt {\gamma^4 \sigma^2 (k-\beta k_x)^2+\gamma^2 r^2}$
and  $k=\sqrt {k_\perp^2 +k_x^2}$.\\
Some examples of the numerical evaluation of this result are shown in Fig.~\ref{fig:fig1}. The parameters are taken to simulate a reasonable experimental setting in fused silica, i.e. $n=1.5$, $L=5$ cm. The refractive index 
perturbation is assumed to be generated via the nonlinear Kerr effect, i.e. $n=n_0+n_2 I$. In fused silica
 the nonlinear Kerr index is $n_2\simeq3\times10^{-16}$ cm$^2$/W \cite{desalvo} and intensities of the order of 
$1-30\times10^{12}$ W/cm$^2$ are achievable within filaments \cite{couairon}. We thus take $\eta=n_2 I=10^{-2}$. We choose two different values of the perturbation radius, $\sigma=1$ $\mu$m [Figs.~\ref{fig:fig1}(a), (b) and (c)] and $\sigma=2$ $\mu$m [Figs.~\ref{fig:fig1}(d), (e) and (f)] and we then vary the degree of superluminality from slightly superluminal to strongly superluminal $\beta=1.1$, 2.1 and 5.1, as indicated in the figure. A first observation is that for small superluminality, the photon emission is peaked at a wavelength, $\lambda_\textrm{max}$,  that scales with the perturbation diameter. Roughly speaking, the maximum emission wavelength is such that $\lambda_\textrm{max}\sim3\sigma$. We also note that, differently to the Cerenkov effect, emission occurs for all angles \emph{inside} the cone, and decays exponentially outside (other differences with respect to the Cerenkov effect are detailed below). 
As the perturbation velocity, or equivalently $\beta$, increases the emission cone angle increases and $\lambda_\textrm{max}$ decreases. In all cases, emission occurs in the near- and mid-infrared wavelength region with maximum photon counts of the order of $10^{-4}-10^{-2}$ thus implying, accounting also for detection for example over a 20 deg angle, $\sim$0.1-10 counts/second with a kHz-repetition rate laser. These numbers are within the limits of commercial, mid-infrared detectors. \\

\noindent \underline{\emph{Correlated pair emission.}} We may also consider measurements on correlated pairs of photons, 
which would represent a robust corroboration that the observed phenomenon has a quantum origin.  \\
 Some qualitative aspects of the pair emission can be obtained from the evaluation of the $\delta$ function in Eq.~(\ref{Nkmu}). On account of this constraint, for a fixed photon emitted with momentum $\vec{k}$, the momentum $\vec{k}'$ of the correlated photon satisfies 
\begin{equation}\label{correlations}
k_x'v-\omega_{\vec{k}'}=-(k_xv-\omega_{\vec{k}}).
\end{equation}
 If we consider the ``superluminality'' cone delimited by the angle $\theta_0=\arccos(\frac{1}{\beta})$, then Eq.~(\ref{correlations}) implies that for every photon {\emph{inside}} the cone, there is a correlated photon {\emph{outside}} the cone. In other words, the photon density for $\beta\sim1$ [see Figs.~\ref{fig:fig1}(a), (d)]  at large angles is extremely small, but not identically zero. As the cone angle increases, i.e. for increasing $\beta$, the photon density increases significantly in the outer region [see Figs.~\ref{fig:fig1}(c), (f)], in keeping with the overall increase of the photon density inside the cone and of the smaller solid angle spanned by the outer region.\\

\noindent \underline{\sl Comparison with other photon production phenomena.} In order to appreciate the novelty and peculiarities of the present photon production mechanism, we give an explicit comparison, highlighting similarities and differences, with other known and possibly competing photon emission phenomena.\\
Cerenkov emission: The condition of superluminality $v>c/n$ required to observe photon emission introduces a link to other superluminal propagation phenomena such as the Cerenkov effect \cite{ginz-book,ginzburg}. However, there are important differences that render the two effects different. The Cerenkov effect requires an electric charge or a dipole moving superluminally in an insulator. The moving charge will induce a re-orientation of the medium dipoles and the subsequent emission of light. Moreover, this effect may be completely described classically and simple trigonometry gives an emission angle for the radiation, $\cos\theta=v/(cn)$. Contrary to these known features, the present photon emission mechanism relies on the propagation of a {\emph{neutral}} refractive index perturbation, i.e. there are no dipoles or charges of any kind in movement. Moreover, there is no clear way to describe this mechanism in terms of classical physics: this is an exquisitely quantum manifestation characterized by the emission of quantum correlated photon pairs. Finally, the photon pairs are not emitted at a specific angle, but one photon of the pair is emitted inside the cone and the other photon of the pair is emitted outside the cone. Therefore, for small $\beta$, one half of the photons is emitted in a small range of angles giving the maximum of emission which can be seen in the Fig.~\ref{fig:fig1} for $\beta=1.1$ and $\beta=2.1$.\\
Four Wave mixing: Intense laser pulses may excite photons from the vacuum via parametric processes such as Four-Wave-Mixing (FWM) in media with third order nonlinearity (also called Kerr media) or Parametric-Down-Conversion (PDC) in media with second order nonlinearity (known as $\chi^{(2)}$ media). The distinguishing feature of these mechanisms is the parametric interaction between the pump and the generated photons and they are thus limited by the so-called phase-matching constraints, i.e. energy and momentum conservation. Thus, in Kerr media, the two emitted photons may in general have different frequencies and the phase velocity of the sum of the emitted fields is determined by the pump pulse {\emph{phase}} velocity. This constrains the emitted photons to a well-defined cone angle and is clearly very different from the described  emission of equal-frequency photons that is determined only by the condition on the pump pulse {\emph{group}} velocity, $v<c/n$. \\
Dynamical Casimir effect: the DCE refers to the emission of photons excited from the vacuum by a mirror (or similar physical effect) moving with a non-uniform acceleration \cite{davies}. Although the refractive index perturbation described here may be assimilated to a moving mirror, it is assumed to be moving with a {\emph{constant}} speed. Moreover the DCE does not present any sort of cut-off or threshold on the mirror velocity. \\
We therefore conclude that the physical effect we are describing is novel and distinct from other known, yet similar, effects. All of these, along with others such as Hawking radiation, may be somehow gathered under the same generic name of quantum friction \cite{davies2}, of which the mechanism proposed here represents a novel realization that holds promise for experimental detection. In this sense, similarly to other superluminal phenomena \cite{ginz-book}, we note that the emitted photons will drain energy from the moving refractive index perturbation which in turn is fed by the pump laser pulse, thus  either reducing its energy until the perturbation is switched off or, alternatively, slowing it down until the superluminality condition is no longer met within the transparency range of the medium. In any case we underline that the presence of dissipation does not modify the results: the evaluation of the emitted photon rate in the realistic experimental setting described above predicts a loss less than a photon-per-pulse. If we consider that the pump pulse will typically have an energy in the $\mu$J or mJ range, this implies that the quantum friction process is far too weak to significantly perturb the pump pulse propagation over short propagation distances.\\ 
\indent In conclusion we have described a novel photon production mechanism that has a two-fold importance: on the one hand  it represents a completely novel  nonlinear optical effect by which an intense Gaussian pulse propagating in a Kerr medium emits correlated photons nearly isotropically within a limited cone angle and with a tunable spectral maximum in the mid-infrared region. The precise correlation properties of the emitted photons are somewhat involved and will be presented in future work. On the other hand, the present effect  has common features with a whole family of vacuum excitation mechanisms, generically referred to as ``quantum friction''  \cite{davies} and that still await experimental observation. \\

The authors gratefully acknowledge discussions with I. Carusotto.



\begin{thebibliography}{99}
\newcommand{\enquote}[1]{``#1''}
\expandafter\ifx\csname url\endcsname\relax
  \def\url#1{{#1}}\fi
\expandafter\ifx\csname
urlprefix\endcsname\relax\def\urlprefix{}\fi

\bibitem{schwinger} 
J.~Schwinger,  Phys. Rev. {\bf 82}, 664 (1951).

\bibitem{hawking}
S.W.~Hawking, Commun. Math. Phys. {\bf 43}, 199 (1975).




\bibitem{philbin}
T.G. Philbin, C. Kuklewicz, S. Robertson, S. Hill, F. K\"onig, and U. Leonhardt, Science, {\bf 319}, 1367 (2008).

\bibitem{faccio}
D. Faccio, , S. Cacciatori, V. Gorini, V.G. Sala, A. Averchi, A. Lotti, M. Kolesik, and J.V. Moloney, arXiv:0905.4426v1 (2009).

\bibitem{milchberg}
I. Alexeev, K.Y. Kim, and H. M. Milchberg, Phys. Rev. Lett., {\bf{88} }, 073901 (2002).

\bibitem{bonaretti}
F. Bonaretti, D. Faccio, M. Clerici, J. Biegert, and P. Di Trapani, Opt. Express {\bf{17}} 9804 (2009).

\bibitem{saari_bessel}
P. Bowlan, H. Valtna-Lukner, M. L\~ohmus, P. Piksarv, P. Saari, and R. Trebino, Opt. Lett. {\bf{34}}, 2276 (2009).

\bibitem{air-fil}
D. Faccio, A. Averchi, A. Lotti, P. Di Trapani, A. Couairon, D. Papazoglou, and S. Tzortzakis, Opt. Express {\bf{16}}, 1565 (2008).

\bibitem{russian-OC}
I. Blonskyi, V. Kadan, O. Shpotyuk, I. Dmitruk, Opt. Commun. {\bf{282}}, 1913 (2009).

\bibitem{ginz-book}
\newblock{V.L.~Ginzburg, {\sl Applications of Electrodynamics in Theoretical
Physics and Astrophysics}. Gordon and Breach Science Publishers, New York (1989).}

\bibitem{ginzburg}
\newblock{V.L.~Ginzburg, 
Sov. Phys. Usp. {\bf 15}, 184 (1972).}



\bibitem{soff}
R.~Sch\"{u}tzhold, G.~Plunien and G.~Soff,
Phys. Rev. A {\bf 58}, 1783 (1998).

\bibitem{belgiornoPRL}
M. Visser, S. Liberati, F. Belgiorno, and D.W. Sciama,
Phys. Rev. Lett. {\bf 83}, 678 (1999).

\bibitem{desalvo}
R. DeSalvo, A.A. Said, D.J. Hagan, E.W. {Van Stryland}, M. Sheik-Bahae, IEEE J. Quant. Electron. {\bf 32}, 1324 (1996)

\bibitem{couairon}
A. Couairon, A. Mysyrowicz, Phys. Rep. {\bf{441}}, 47 (2007).





\bibitem{davies}
S.A. Fulling, P.C.W Davies, Proc. R. Soc. London A, {\bf 348}, 393 (1976).

\bibitem{davies2}
P.C.W. Davies, J. Opt. B: Quantum Semiclass. Opt., {\bf 7} S40 (2005). 





\end{thebibliography}
\end{document}